\documentclass[aps,prl,groupedaddress,twocolumn,floatfix]{revtex4}
\usepackage{graphicx,amssymb,amsmath,amsbsy}
\usepackage[T1]{fontenc}
\usepackage{natbib} 
\def\slantfrac#1#2{\kern.1em^{#1}\kern-.3em/\kern-.1em_{#2}}
\def\b#1{\mathbf{#1}}
\def\bs#1{\boldsymbol{#1}}
\def\m#1{\mathrm{#1}}

\begin{document}
\pagestyle{empty}
\medskip
{\Large{\centerline{The Sagnac Effect in Superfluids}}}
\bigskip
\begin{center}
 E. Varoquaux\\
     CEA-Saclay/DSM/DRECAM/SPEC \\    
    91191 Gif-sur-Yvette, France \\ 
 and\\
 G. Varoquaux\\ 
    Laboratoire Charles Fabry de l'Institut d'Optique\\
    UMR 8501 du CNRS\\
    Campus Polytechnique, RD128 -- 91129 Palaiseau, France
\end{center}
\bigskip

In this Letter, we wish to consider some problems of interpretation of
Sagnac-type experiments with beams of cold atoms and with superfluids. In
particular, we have in mind to disprove Malykin's following comment on the
latter system made in his otherwise well-documented and comprehensive review
article on the Sagnac effect \cite{Malykin:00}:
 
``{\it It should be noted that the inertial properties of waves (or
wave packets, for that matter) are made use of in such
gyroscopic instruments as solid-state wave gyroscopes
and also gyroscopes whose principle of action is grounded on
the macroscopic quantum properties of superfluid helium.
These instruments along with the Foucault
pendulum and mechanical gyroscopes are
applied to determine the angular position in space. In
contrast, devices in which the Sagnac effect provides the
working principle /.../ serve as angular velocity pickups. This
makes the fundamental distinction between instruments
based on the Sagnac effect and those in which the property
of physical bodies or wave packets to maintain orientation in
space is employed.}''

In spite of the maturity of its subject matter, Malykin's review stirred the
need for further clarification and comments \cite{Malykin:02,Rizzi:04}. Here,
we want to point out that, contrarily to the statement above, superfluid
interferometers do measure the absolute angular velocities of the platforms on
which they are mounted. But, more importantly, we also attempt to address the
somewhat widespread (see {\it e.g.}  \cite{Malykin:00,Nandi:02,Rizzi:04})
misconception that superfluid rotation sensors, unlike atomic beams gyros,
would not belong to the same class of quantum interference effects as the
Sagnac light-wave experiments.

The Sagnac effect is no longer an object of sole academic curiosity studied to
prove (or disprove in the eyes of some, Sagnac being one) the foundations of the
theory of relativity; it has spread to applications of daily usefulness such
as the ring laser gyros in inertial guidance devices and the Global
Positioning System.

For these purposes, the effect is well understood
\cite{Landau:ClassicalFields,Chow:85,Stedman:97,Ashby:03}.  In the classic
textbook of Landau and Lifschitz \cite{Landau:ClassicalFields} the rotating
frame of reference, embodied by orbiting satellites carrying atomic clocks,
our Earth, or turntables supporting interferometers, is treated as an
accelerated frame from the point of view of general relativity. In such frames,
characterised by a space-time metric $ -\m d s^2= g_{00}\m d (x^0)^2 +
2g_{0i}\m dx^0 \m dx^i+g_{ii}\m d(x^i)^2$, clocks can be synchronised for
infinitely close points by time shift $\m d t = -g_{0i} \m dx^i/g_{00}$. If a
clock is transported around a finite path $\mathit\Gamma$ in a frame rotating
with velocity $\b\Omega$, the resulting total time shift is
(\cite{Landau:ClassicalFields} \S 89)
\begin{equation}        \label{TimeDifference} 
  \Delta t = \frac{1}{c}\,\oint_{\mathit\Gamma} \frac{g_{0i}\m d x^i}
    {-g_{00}} 
    =\oint_{\mathit\Gamma} \frac{\b\Omega \!\bs\times\!  \b r \bs\cdot\m d\b r}
    {c^2-(\b\Omega\!\bs\times\!\b r)^2}
    \simeq \frac{2}{c^2}\, \b\Omega \bs\cdot \b S  
\end{equation}
$\b S$ being the vector area subtended by $\mathit\Gamma$. Time delay
(\ref{TimeDifference}) between the reading of the transported clock and that
of the clock standing still on the rotating platform lies at the root of the
Sagnac effect. Such a point of view has been held long ago by Langevin
\cite{Langevin:21-37} and others \cite{Malykin:00}.

For light waves with angular frequency $\omega$, the corresponding phase shift
reads
\begin{equation}        \label{WavePhaseShift}
  \Delta \varphi = \omega\Delta t = \frac{4\pi\, \b\Omega \bs\cdot \b S }{\lambda c} 
\end{equation}
where $\lambda$ is the wavelength in vacuum, $\lambda = 2 \pi c /\omega$. 

Formulae (\ref{TimeDifference}) and (\ref{WavePhaseShift}) are usually derived
for optical interferometric experiments in the framework of the special theory
of relativity, using Lorentz boosts to calculate to velocity of the moving
clock or wave (see {\it e.g.}  \cite{Malykin:00,Rizzi:04}). Since Sagnac's
early experiments in 1913, their validity has been confirmed in detail with
optical interferometers and by atomic clock transportation as reviewed for
instance in \cite{Post:67,Malykin:97}.
 
New physical systems, to which the same conceptual framework as for the
original Sagnac experiment can be applied, have been studied in the past
twenty years or so when it became possible to split beams of particles and to
have them recombine and interfere.  Interferometers were built using neutrons
and electrons, and, more recently, atomic beams and superfluids.  Together
with these experimental advances came alternative interpretations of the
effect. 

Let us deal first with particles -- electrons, neutrons or atoms --
represented by localised wavepackets with a slowly-varying overall phase
$\varphi$. These wavepackets can be treated in a quasi-classical approach: the
phase is related to the classical action $\varphi = {\cal S/\hbar}$. This
action can be computed in a rotating frame following for instance
\cite{Hasselbach:93,Storey:94}.  The Lagrangian for a free particle with mass
$m$ located at position $\b r$ and moving with velocity $\b v$ in a reference
frame rotating with angular velocity $\b \Omega$ is expressed by:
\begin{equation}        \label{Lagrangian}
  {\cal L}(\b r,\b v) = \frac{m}{2} v^2 + m\, \b\Omega\bs\cdot(\b r\!\bs\times\!\b v)
    + \frac{m}{2}(\b\Omega\!\bs\times\!\b r)^2 \; .
\end{equation}

The discussion is restricted to the case of slow rotations, which are treated
as a small perturbation. The action is then obtained as the integral of the
Lagrangian, Eq.(\ref{Lagrangian}), over the unperturbed path of the particle,
along which its velocity $v$ is constant. To first order in $\Omega r/c$, the
last term in Eq.(\ref{Lagrangian}) can be neglected and the expression of the
action reduces to
\begin{equation}        \label{Action}
    {\cal S} = \int_\mathit\Gamma  \! \m d t \, {\cal L}\Big(\b r(t),\b
    v(t)\Big) 
    =  m \b\Omega\,\bs\cdot\int_\mathit\Gamma \! \m d t\,[\b r(t)\!\bs\times\!\b v(t)]  \; .
\end{equation}
Since $\b v(t) = \m d \b r(t)/\m d t$, the last integral in Eq.(\ref{Action}) is
twice the area swept along $\mathit\Gamma$. For a closed path, the change of
the phase of a wavepacket upon completing a round trip involves the area $\b
S$ subtended by $\mathit\Gamma$:
\begin{equation}        \label{ParticlePhaseShift}
  \Delta\varphi = \frac{m}{\hbar}\, \b\Omega\,\bs\cdot\oint_\mathit\Gamma \b
    r\!\bs\times\!\m d\b r = \frac{m}{\hbar} \, 2\,{\b\Omega\bs\cdot\b S}   \; .
\end{equation}
Equation (\ref{ParticlePhaseShift}) expresses the Sagnac phase shift for
massive particles as obtained from a purely non-relativistic kinematical approach.

We now turn to the helium liquids. The inertial properties of superfluids have
been the subject of numerous studies \cite{Leggett:99}. They are governed by
the existence of an order parameter that acts as a macroscopic wavefunction
with a well-defined overall phase $\varphi$. 
The superflow velocity is proportional to the gradient of this
phase,
\begin{equation}        \label{SuperfluidVelocity}
{\bf v}_{\m s} = (\hbar/m)\nabla\varphi \; ,
\end{equation}
where $m$ is the atomic mass, $m_4$, for $^4$He and the Cooper pair mass,
$2m_3$ for $^3$He-B \cite{AnisotropicPhase}.  No gauge field added to
$\varphi$ can allow this expression to transform through rotation of the
reference frame; it only holds in inertial reference frames.  

For a pool of superfluid in the shape of a torus, the continuity of the
phase requires the circulation of the velocity along a closed contour
$\mathit\Gamma$ threading the torus to be quantised {\it in the inertial
frame} \cite{PhaseBias}:
\begin{equation}        \label{VelocityCirculation}
\oint_{\mathit\Gamma} {\bf v}_s \bs\cdot{\m d}{\bf r}=
\frac{\hbar}{m}\,\oint_{\mathit\Gamma}\,\nabla\varphi \bs\cdot{\m d}{\bf
r} =  n\kappa   \; , \end{equation}
where $\kappa = 2\pi\hbar/m$ is the quantum of circulation and $n$ an
integer.
 
This quantum feature of superfluids has been demonstrated experimentally by
setting the toroidal vessel into rotation.  As shown by Hess and Fairbank
\cite{Hess:67}, states of circulation quantised in the inertial frame
spontaneously appear at the superfluid transition. In particular, a state of
zero circulation, $n=0$, the so-called Landau state, can exist. The superfluid
fully decouples from its container: it settles at rest with respect to the
distant stars, that is, in motion with respect to the container walls.

At finite temperature, a non-superfluid fraction appears in the fluid, formed
by the thermally-excited elementary excitations in the superfluid, the phonons
and rotons for $^4$He, thermal quasi-particles and quasi-holes for $^3$He.  As
shown by Reppy and Lane \cite{Reppy:65}, the superfluid velocity circulation,
defined by Eq.(\ref{VelocityCirculation}), is the conserved quantity as the
temperature, hence the superfluid fraction, changes, not the angular
momentum associated with the motion of the superfluid component.

A rotating superfluid is not simply a classical inviscid fluid with angular
momentum; circulation quantisation constitutes a stricter constraint, immune
to perturbations by moving boundaries and to temperature changes, as
illustrated by the experiments mentioned above and many others. These properties
fundamentally follow from Eq.(\ref{SuperfluidVelocity}) and the continuity of
the quantum phase throughout the superfluid. They entail the existence of a
Sagnac effect.

In a frame rotating with absolute rotation ${\b \Omega}$ the superfluid
velocity transforms according to ${\b v}'_\m s ={\b v}_\m s - {\b \Omega}
\!\bs\times\! {\bf r}$ and the quantisation of circulation condition
(\ref{VelocityCirculation}) reads
\begin{equation}        \label{Circulation}
    \oint_{\mathit\Gamma}{\bf v}'_s\bs\cdot{\m d}{\bf r} 
    =\oint_{\mathit\Gamma}({\bf v}_s-\b\Omega\!\bs\times\!\b r) \bs\cdot{\m d}{\bf r}
    =  n\kappa - 2\,{\bf \Omega}\bs\cdot{\bf S} \;. 
\end{equation}
The last term to the right of Eq.\,(\ref{Circulation}) amounts to a
non-quantised contribution to the circulation in the rotating frame that
varies with rotation vector ${\bf\Omega}$. This circulation gives rise to a
phase change $\Delta\varphi= (m/\hbar)\,2\,{\bf{\Omega\bs\cdot S}}$ that,
measured by means of Josephson-type devices \cite{Avenel:96-97}, gives access
to the rotation vector ${\b \Omega}$, contrarily to the statement in
\cite{Malykin:00} quoted above. The superfluid gyros in \cite{Avenel:96-97}
are gyrometers, not gyroscopes.
 
The phase difference stemming from Eq.(\ref{Circulation}) is precisely that
arising from the Sagnac effect for particles with mass $m$,
Eq.(\ref{ParticlePhaseShift}). This coincidence is not simply formal: an
applied rotation has the same effect on the phase of an atomic wavepacket in
an atom-interferometric experiment than on that of the superfluid macroscopic
wavefunction in a toroidal vessel.

If we now invoke wave-particle duality and introduce the de Broglie wavelength
of the particle of mass $m$ and velocity $v$, namely $\lambda_\m B =
2\pi\hbar/(m\,v)$, in Eq.(\ref{ParticlePhaseShift}), we find
\begin{equation}        \label{QuantumPhaseShift}
  \Delta \varphi = \frac{4\pi\, \b\Omega \bs\cdot \b S }{\lambda_\m B v} \; .
\end{equation}
For photons in vacuum, $v=c$, and we recover Eq.(\ref{WavePhaseShift}). 

In a rotating material medium such as a glass fibre ring gyro, the simple
Eq.(\ref{WavePhaseShift}) does not hold. It is necessary to consider both the
wave propagating in the corotating direction and that in the counterrotating
direction to eliminate the refraction properties of the medium (see {\it e.g.}
\cite{Arditty:81} for a discussion).  This circumstance takes advantage of the
reciprocity principle to cancel out the retarded propagation of the light
signals in opposite directions along precisely the same travel path. What is
left is the difference in clock readings, Eq.(\ref{TimeDifference}).

Other examples of the same kind of cancellation between counterrotating waves
are discussed by Malykin \cite{Malykin:00} (see also \cite{Rizzi:04}). For
interferometry with massive particles, the beam-deflecting devices acting as
``mirrors'' introduce additional phase shifts that must be taken into account.
So do gravity and electromagnetic fields. Each separate experiment requires
special considerations (see \cite{Neutze:98} for electrons,
\cite{Borde:01-03} for atoms). In most instances, Eqs.(\ref{TimeDifference})
and (\ref{QuantumPhaseShift}) for the Sagnac effect are found to be obeyed.

Let us emphasise that all massive particle interferometric experiments obey
Eq.(\ref{QuantumPhaseShift}) and belong to the same class. The superfluid is
not the odd man out. It offers so far the only experimental situation in which
a matter-wave field, coherent over the full length of a pickup loop, is
involved but it is quite conceivable that, in a near future, Sagnac-type
experiments will be conducted with Bose-Einstein condensates of ultra-cold
atoms. The required techniques are on the verge of becoming available
\cite{Bongs:04}. An atom of a given atomic species could be made to interfere
with itself or collectively.  In the first case - atomic beam experiments -
each single atom interferes with itself after having travelled along either
arms of the rotating interferometer. In the second - Bose-condensed - case,
the condensate sits nearly idle between the two banks of a Josephson junction,
where it interferes with itself. The interference pattern arises from the
overlap of the macroscopic wavefunction on one side of the junction with the
weakly coupled part that leaks out from the other side.  There is no breach of
conceptual continuity between superfluid and particle Sagnac experiments: we
have, on the one hand, all massive particles, matter waves, on the other,
light signals, clocks and photons.

The Sagnac phase shift for massive particles, Eq.(\ref{QuantumPhaseShift}),
has been amply verified by classic experiments on electronic Cooper pairs
\cite{Zimmerman:65}, neutrons \cite{Werner:79-94,Staudenmann:80}, electrons in
vacuum \cite{Hasselbach:93,Neutze:98}, and atom beams
\cite{Riehle:91,Lenef:97,Gustavson:97-00}. For superfluid helium, the same
equation underpins the experiments reported in
\cite{Avenel:96-97,Schwab:97-Bruckner:03,Mukharsky:00-04-Avenel:04}.  Yet,
this equation differs markedly from Eq.(\ref{WavePhaseShift}), quantitatively,
by a factor $mc^2/\hbar\omega \sim 10^{10}$ to $10^{11}$, and qualitatively,
because Einstein's relativity does not enter its derivation.

There are several, equivalent, ways to restore explicit relativistic
invariance for massive particles and superfluids.

It is possible \cite{Anandan:03,Hendricks:90} to derive {\it a priori} the
rotation terms appearing in Lagrangian (\ref{Lagrangian}) from a fully
general-relativistic description of the matter-wave field. The particle
quantum field is solution of a Dirac-like equation (or Proca, or higher
order). In the rotating frame, the curved metric appears through the Dirac
gamma matrices, and their low velocity expansion yields a Hamiltonian and,
correspondingly, a Lagrangian that generalises (\ref{Lagrangian}). Rotation
terms in this Lagrangian are a direct expression of the effects of the local
space-time curvature on the phase of the quantum field; the Sagnac term for
light waves has the same physical origin.

In the relativistic form of the Lagrangian for weakly-interacting particles,
the kinetic energy term in Eq.(\ref{Lagrangian}) is replaced by
$-mc^2\,(1-v^2/c^2)^{1/2}$ (see \cite{Anandan:81,Riehle:91,Neutze:98}).  A
frequency such that $\hbar\omega = mc^2$ appears that turns
Eq.(\ref{WavePhaseShift}) formally into Eq.(\ref{QuantumPhaseShift}). Massive
and massless particles are thus put on the same footing. This prescription has
been re-examined recently on different grounds by a number of authors for
massive particles \cite{Ryder:01,Nandi:02,Rizzi:04} and by Volovik for
superfluid helium \cite{Volovik:03}.




For superfluids, we can take a more direct approach. A relativistic two-fluid
model can be built over the usual Landau superfluid hydrodynamics by imposing
Lorentz invariance as done in \cite{Lebedev:82-Carter:92}.
The invariant velocity circulation, the generalisation of
Eq.(\ref{VelocityCirculation}), reads
\begin{equation}        \label{CovariantCirculation}
  \int_\Xi\{ v'_0 \m dx^0 + v'_i \m dx^i\} = n\kappa \; ,
\end{equation}
where $(v'_o,v'_i)$ is the four-velocity in the rotating frame $(c^2+\b v'_\m
n\bs\cdot \b v'_\m s, -\b v'_\m s)$. Both the normal fluid velocity $v'_\m n$
and the superfluid velocity $v'_\m s$ are small compared to $c$ so that the
time-like component of the four-velocity reduces to $c^2$. The integration
over $\Xi$ is an actual loop integral only for the space-like components. The
corresponding world line is not closed because the time for synchronised
clocks varies as \mbox{$\m dx^0 = -g_{0i} \m dx^i/g_{00}$}. Making use of
Eq.(\ref{TimeDifference}), we recover Eq.(\ref{Circulation}),
\begin{equation}        \label{Circulation2}
  \oint_\Gamma v'_i \m dx^i = n\kappa + \int\! c^2 g_{0i} \m  dx^i/g_{00} =
  n\kappa - \frac{2}{c^2}\, \b\Omega
  \bs\cdot \b S  \; , 
\end{equation}
which establishes a unifying link between superfluid
physics and the relativistic particle approach.  It shows that the
effect described by Eqs.(\ref{WavePhaseShift}) and (\ref{QuantumPhaseShift})
is one and the same in spite of the quantitative and qualitative differences
stated above. 

Thus, Einstein-synchronised clocks provide the time standard by which phase
differences can be kept track of in all the studied physical systems. As
appropriately summarised by D.M. Greenberger \cite{Greenberger:83}, Sec. IX,
for neutron interferometry experiments: ``{\it the phase shift} (in the
rotating interferometer) {\it is seen to be caused by the different rates at
  which a clock ticks along each of the two beams}''.

Needless to say, low temperature experiments, and even those in cold-atom or
neutron physics, in no way measure relativistic corrections to
Eq.(\ref{QuantumPhaseShift}) derived for massive particles. The experimental
implications of the observation of the Sagnac phase shifts are that no
reference to special or general relativity need be made. In fact, the
derivation of Eq.(\ref{QuantumPhaseShift}) makes no explicit reference to
Einstein's relativity. The non-relativistic limit, obtained by letting
$c\rightarrow\infty$, leaves Eq.(\ref{WavePhaseShift}) for the phase shift
unchanged. Clocks and light-wave experiments, which involve no rest mass
energy, are, for their part, fully relativistic. The reference to clocks tied
to a particle rest energy provides a fully covariant formalism to describe the
Sagnac effect; it bears no direct relevance to laboratory observations but
provides a common viewpoint on the various physical systems.

We hope to have clarified the case for Sagnac experiments in superfluids. As
those with atoms, neutrons, and electrons, they do obey
Eq.(\ref{TimeDifference}) when the proper transcription to the time domain is
effected.  They share with clock transportation the feature that the relevant
variables, superfluid phase or clock time, are defined and obey Eqs.
(\ref{TimeDifference}) and (\ref{QuantumPhaseShift}) along any given path,
irrespective of the details of the paths of well-balanced interferometric
devices. Also, they demonstrate a notably extreme case of ``giant matter
waves'', close to the borderline between quantum systems and classical ideal
fluids but resting on the existence of a quantum phase, which is a
prerequisite for the appearance of phase shifts, circulation quantisation, and
Josephson interference patterns.

Thus, to summarise: (1)~The Sagnac effect takes a particularly simple form in
superfluids as the order parameter phase is a macroscopically defined and
directly measurable quantity
\cite{Avenel:96-97,Schwab:97-Bruckner:03,Mukharsky:00-04-Avenel:04}; (2)~Its
experimental implementation varies considerably between various physical
systems but a unifying, relativistic, formalism is offered by clock
transportation -- massive quantum particles, superfluids, waves, and actual
clocks all carrying their own time reference, as implied before by a number of
authors ({\it e.g.}  \cite{Anandan:81,Greenberger:83,Dieks:90}).

We gratefully acknowledge informative discussions with Alain Comtet, Thierry
Jolic{\oe}ur, Tony Leggett, Lev Pitaevski, and useful comments from Pertti
Hakonen. 


\begin{thebibliography}{39}
\bibitem[{Ani()}]{AnisotropicPhase}
\bibinfo{note}{The case of the anisotropic phase $^3$He-A is more complicated
  because gradients of $\b l$, the orbital anisotropy axis, also contribute to
  the right hand side of Eq.(\ref{SuperfluidVelocity}).}
\end{thebibliography}
\newcommand{\noopsort}[1]{}

\end{document}